\documentclass[12pt,longbibliography]{article}
\usepackage{graphicx}
\usepackage{caption}
\usepackage{subcaption}
\usepackage{amsmath}
\usepackage{xspace}
\usepackage[comma,square,numbers,sort&compress]{natbib}
\RequirePackage{lineno}
\usepackage[pdftex]{hyperref} 
\hypersetup{colorlinks=true, linkcolor=blue, citecolor=blue, urlcolor=black, filecolor=blue}

\newcommand\pubdate{\today}
\newcommand{\polpp}{\text{$p^{\uparrow}$+$p$}\xspace}
\newcommand{\polpA}{\text{$p^{\uparrow}$+$A$}\xspace}
\newcommand{\polpAu}{\text{$p^{\uparrow}$+Au}\xspace}
\newcommand{\polpAl}{\text{$p^{\uparrow}$+Al}\xspace}
\def\Title#1{\begin{center} {\Large #1 } \end{center}}
\def\Author#1{\begin{center}{ \sc #1} \end{center}}
\def\Address#1{\begin{center}{ \it #1} \end{center}}

\newcommand\pubblock{\rightline{\begin{tabular}{l}  \\ 
         \pubdate  \end{tabular}}}
\newenvironment{Abstract}{\begin{quotation}  }{\end{quotation}}
\newenvironment{Presented}{\begin{quotation} \begin{center} 
             PRESENTED AT\end{center}\bigskip 
      \begin{center}\begin{large}}{\end{large}\end{center} \end{quotation}}

\begin{document}
\begin{titlepage}
 \pubblock
\vfill
\Title{Transverse Single Spin Asymmetries of charged hadrons at forward and backward rapidity from $p^{\uparrow}+p$, $p^{\uparrow}+\mathrm{Al}$, and $p^{\uparrow}+\mathrm{Au}$ collisions in PHENIX}
\vfill
\Author{Jeongsu Bok (for the PHENIX collaboration)}
\Address{Pusan National University, Busan, South Korea\\
         New Mexico State University, Las Cruces, USA}
\vfill
\begin{Abstract}
Transverse Single Spin Asymmetries (TSSAs) in transversely polarized proton-proton collisions ($p^{\uparrow}+p$) have been a fruitful source for studying the spin structure of the proton. In the 2015 RHIC data taking periods, collisions of polarized protons with nuclei ($p^{\uparrow}+A$) were made for the first time. The measurements of TSSAs in $p^{\uparrow}+p$ and $p^{\uparrow}+A$ collisions can provide a unique opportunity to investigate the origin of TSSA in a gluon-rich target nucleus and provide a tool to study nuclear effects in $p+A$ collisions. This presentation will report PHENIX results of TSSAs for charged hadrons ($h^{\pm}$) at forward and backward rapidity ($1.4<|\eta|<2.4$) over the transverse momentum ranges $1.25<p_{T}<7.0 \mathrm{\ GeV}/c$ and Feynman-$x$ ranges ($-0.2<x_{F}<0.2$) from $p^{\uparrow}+p$, $p^{\uparrow}+\mathrm{Al}$, and $p^{\uparrow}+\mathrm{Au}$ collisions at $\sqrt{s_{\rm NN}}= 200$ GeV.
\end{Abstract}
\vfill
\begin{Presented}
DIS2023: XXX International Workshop on Deep-Inelastic Scattering and
Related Subjects, \\
Michigan State University, USA, 27-31 March 2023 \\
     \includegraphics[width=9cm]{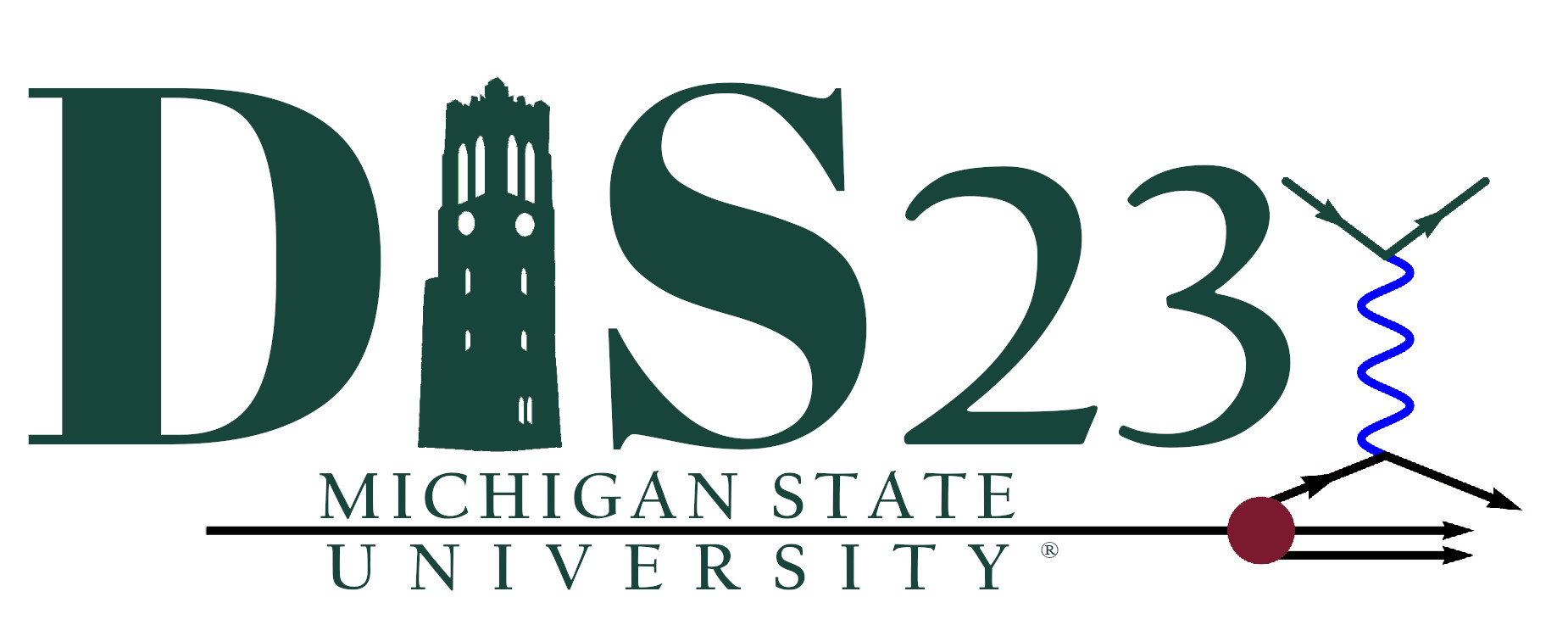}
\end{Presented}
\vfill
\end{titlepage}


\section{Introduction}
Since the 1970's, large Transverse Single Spin Asymmetries (TSSAs) in single hadron production have been observed at forward rapidity in transversely polarized proton-proton collisions ($\polpp \rightarrow h + X$) in a wide range of collision energies. The phenomena have provided an important motivation to investigate spin-momentum correlations inside a nucleon. 
The analyzing power $A_{N}$ is defined as the left-right asymmetry of the hadrons with respect to the polarization of the proton where the direction of polarization is perpendicular to the beam direction.
\begin{equation}
A_{N}(\phi) = \frac{\sigma^{\uparrow}-\sigma^{\downarrow}}{\sigma^{\uparrow}+\sigma^{\downarrow}}    
\end{equation}

\begin{figure}[thb]
\centering
\includegraphics[width=\linewidth]{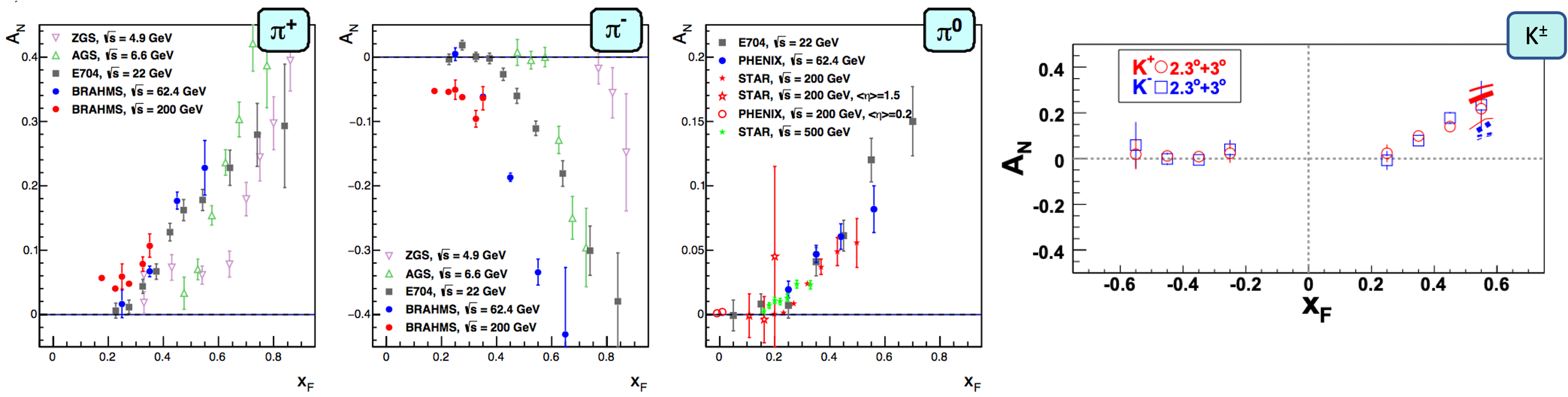}
\caption{\label{fig:AN_collection}
Transverse single spin asymmetry measurements for $\pi^{0}$ and $\pi^{\pm}$ at different center-of-mass energies~\cite{Aschenauer:2016our} and $K^{\pm}$ at RHIC energy as a function of Feynman-x~\cite{BRAHMS:2008doi}.}
\end{figure}

Figure~\ref{fig:AN_collection} shows the measurements of $A_N$ for neutral and charged pions as a function of Feynman-$x$ ($x_F$) at different energies~\cite{Aschenauer:2016our}. The magnitude of $A_N$ increases as a function of $x_F$ at large positive $x_F$ from low to high energies up to 500 GeV. 
$A_N$ for $\pi^{+}$ has a positive sign at $x_F>0$, and $A_N$ for $\pi^{-}$ has comparable size, but opposite sign. On the right side of Fig.~\ref{fig:AN_collection}, charged kaons result at 62.4 GeV~\cite{BRAHMS:2008doi} show that $A_N$ for negatively and positively charged kaons have the same sign at $x_F>0$, which is different from pions. 

There are possible origins to explain these nonzero asymmetries. As an initial state effect, the correlation between the proton spin and the transverse momentum of the parton is called the Sivers mechanism. The final state effect is called the Collins mechanism and describes the correlation between the spin of the fragmenting quark and the transverse momentum of the final state hadron. It is convoluted with the Transversity distribution functions, the correlation between the spin of a proton and the spin of a quark. For the single hadron production in proton-proton collisions, the twist-3 collinear factorization approach is applicable, and the asymmetry arises from a twist-3 multi-parton correlation and a twist-3 fragmentation function.

In addition to the proton-proton collisions, the polarized proton-nucleus collision (\polpA) is proposed to solve this puzzle with an interplay between small-$x$ physics and spin physics. Since the mechanisms are expected to have different nuclear target dependences of $A_N$, measuring $A_N$ in \polpA collisions can help to clarify the origin of the $A_N$~\cite{Kang:2011ni, Kovchegov:2012ga, Hatta:2016wjz, Hatta:2016khv, Benic:2018amn}. Also, nuclear dependence of $A_N$ can be a probe for the saturation scale in the nucleus~\cite{Kang:2011ni}.

\section{Experimental Setup}

\begin{figure}[thb]
\centering
\includegraphics[width=0.7\linewidth]{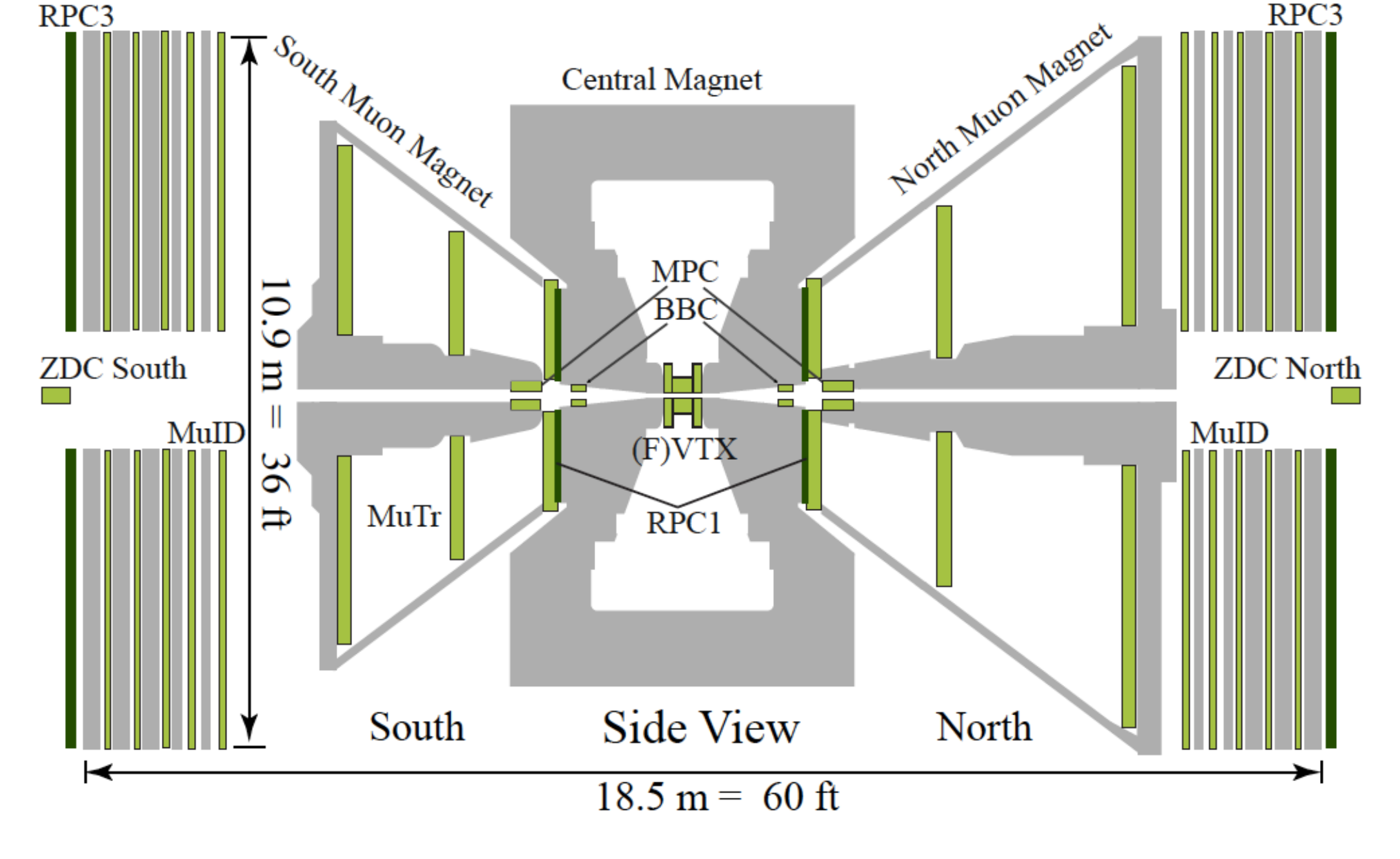}
\caption{Side view of the PHENIX detector in 2015.}
\label{fig:Phenix_2015}
\end{figure}

The Relativistic Heavy Ion Collider (RHIC) can collide various kinds of ions and protons in a wide range of energies, up to 510 GeV for proton-proton collisions and 200 GeV for ion collisions. Also, RHIC is the first and only collider which can collide polarized protons. In addition, the first polarized proton-nucleus collisions were made during the 2015 run period. The PHENIX experiment has two muon arms which cover $-2.2<\eta<-1.2$ and $1.2<\eta<2.4$. In the muon arms, charged hadrons such as pions and kaons stop in the intermediate layer of the muon identifier while muons of large momentum over 3 GeV/$c$ penetrate the whole detector. Therefore reconstructed tracks with a minimum momentum requirement are selected as charged hadrons~\cite{PHENIX:2019gix}. The analysis methods using the muon arm have been shown in previous studies on heavy flavor and charged hadrons~\cite{PHENIX:2017wbv,PHENIX:2019ouo}.

\begin{figure}[thb]
\centering
  \begin{subfigure}[b]{0.4\linewidth}
    \includegraphics[width=\linewidth]{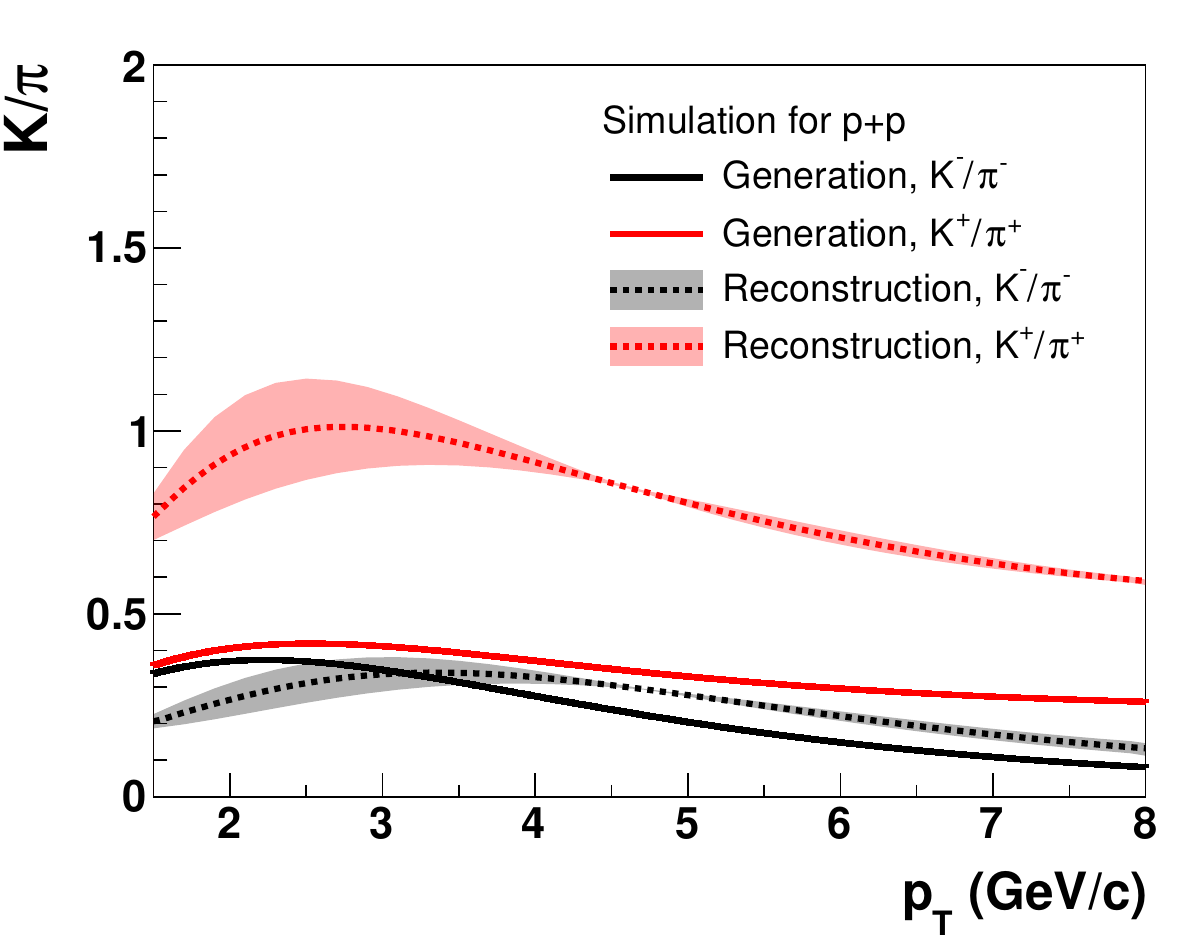}
    \label{fig:Kpi_ratio_pp_pt}
    \caption{$K/\pi$ ratio in $p+p$ collisions.}
  \end{subfigure}
  \begin{subfigure}[b]{0.4\linewidth}
    \includegraphics[width=\linewidth]{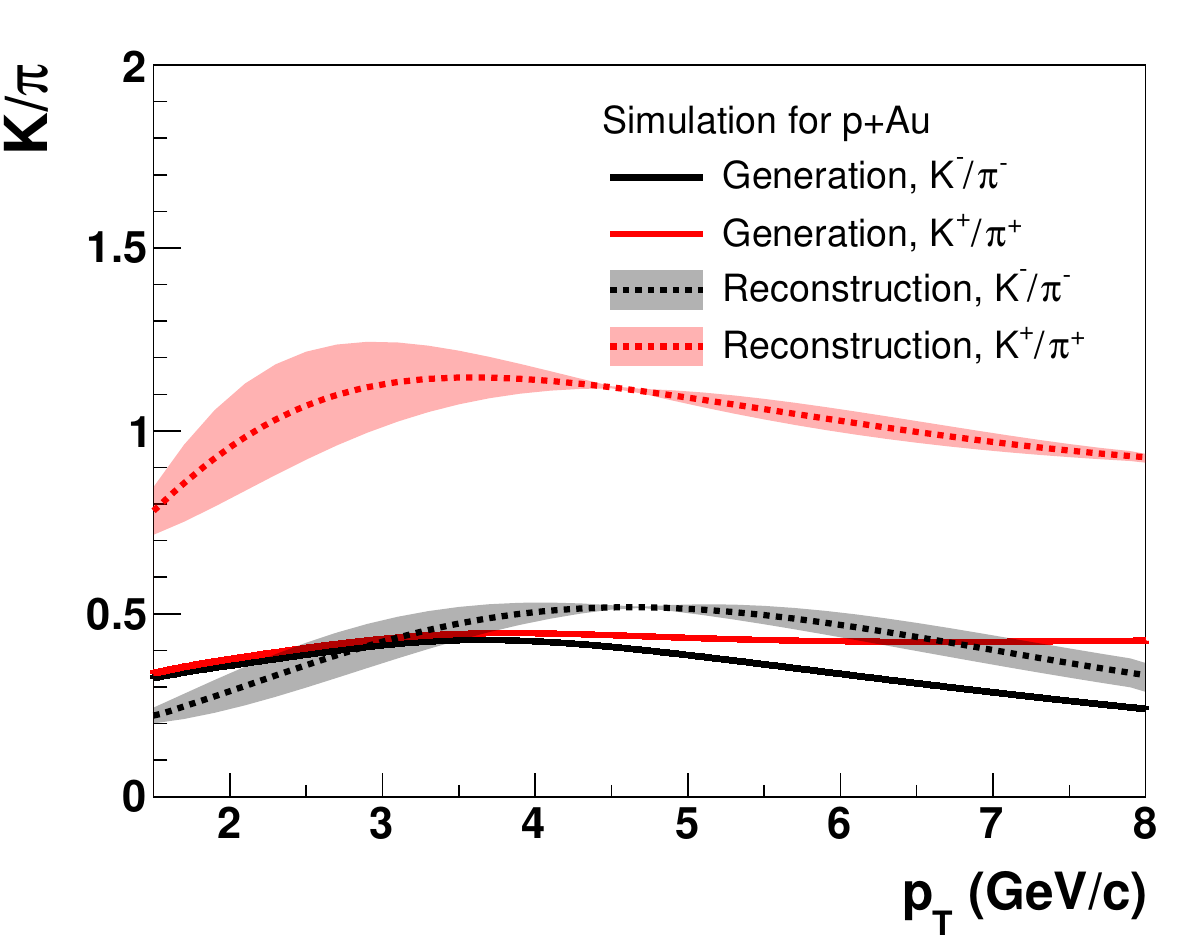}
    \label{fig:Kpi_ratio_pAu_pt}
    \caption{$K/\pi$ ratio in $p+A$ collisions.}
  \end{subfigure}		
\caption{
Estimated $K/\pi$ ratios at the collision vertex (Generation) and $K/\pi$ ratios for reconstructed muon arm tracks in in the {\sc geant4} simulation (Reconstruction) for (a) $p+p$ and (b) $p+A$ collisions~\cite{PHENIX:2023axd}. The variations on the reconstructed $K/\pi$ ratios are from the different physics lists in the GEANT4 simulation.}
\label{fig:Kpi_ratio}
\end{figure}

Based on the PYTHIA and HIJING event generators and GEANT4 simulation, the particle composition in the measured charged hadron sample was estimated~\cite{PHENIX:2012itj,PHENIX:2013txu}. 
Fig.~\ref{fig:Kpi_ratio} shows the estimated $K/\pi$ ratios at the collision vertex (Generation) and in the reconstructed muon arm tracks from $p+p$ and $p+A$ collisions. The variations from the different physics lists in the GEANT4 simulation are considered. Due to interaction with the detector material, the composition of the reconstructed particle is modified from the composition at the collision vertex.

\section{Results and Discussion}

\begin{figure}[thb]
\centering
\begin{subfigure}[b]{0.49\linewidth}
\includegraphics[width=1.0\linewidth]{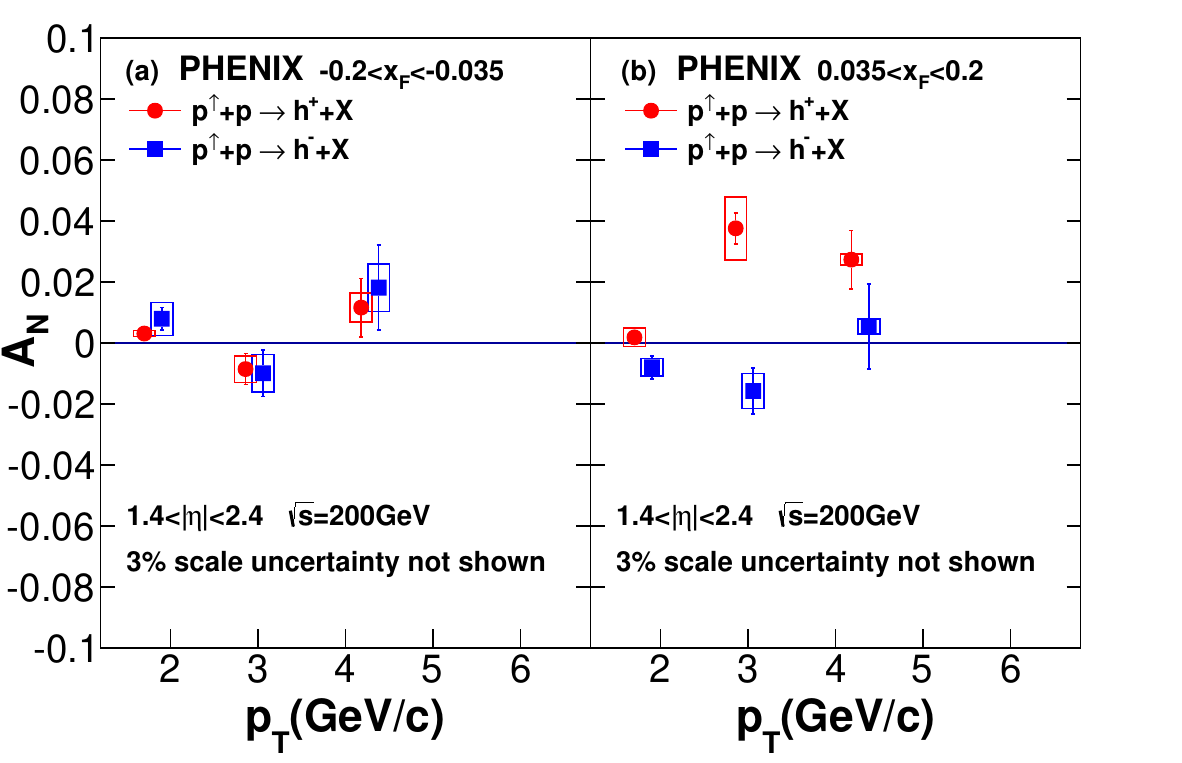}
\label{fig:AN_pt_pp}
\end{subfigure}
\begin{subfigure}[b]{0.49\linewidth}
\includegraphics[width=1.0\linewidth]{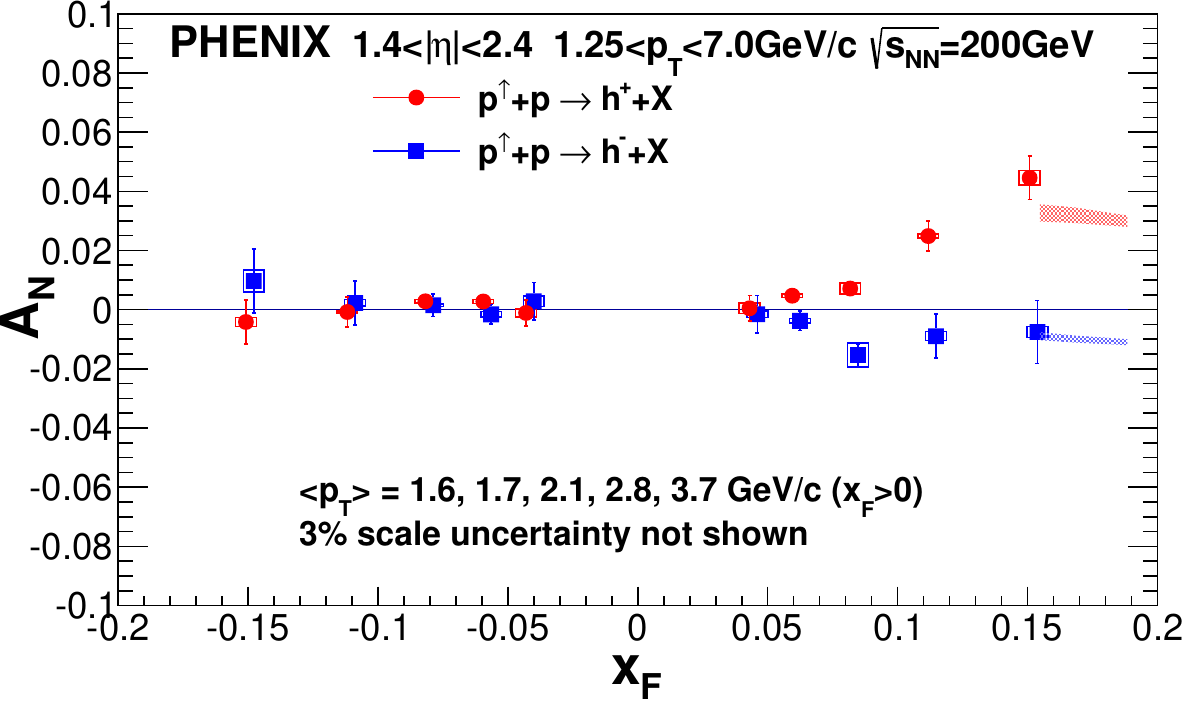}
\label{fig:AN_xf_pp_band}
\end{subfigure}
\caption{\label{fig:AN_pp}
(Left) $A_N$ of charged hadrons from \polpp collisions as a function of $p_T$ in the backward ($x_F<0$) and forward ($x_F>0$) region. 
(Right) $A_N$ of charged hadrons from \polpp collisions as a function of $x_F$~\cite{PHENIX:2023axd}. 
Bands are obtained twist-3 model calculations~\cite{Gamberg:2017gle} by varying the $K/\pi$ ratio by $\pm 30 \%$ relative to the central value.}
\end{figure}

Fig.~\ref{fig:AN_pp} shows $A_N$ for negatively and positively charged hadrons in \polpp collisions as a function of $p_T$ and $x_F$~\cite{PHENIX:2023axd}. $A_N$ at $x_{F}<0$ are consistent with zero for both charges. At $x_{F}>0$, $A_N$ for positively-charged hadrons is positive and increases as a function of $x_F$. Positively charged hadrons are mostly $\pi^{+}$ and $K^{+}$ and they have similar increasing trend as a function of $x_F$ at $x_{F}>0$, the $A_N$ is consistent with it. $A_N$ for negatively charged hadrons at $x_{F}>0$ shows small negative asymmetries. A possible explanation is the partial cancellation between $\pi^{-}$ and $K^{-}$ where the opposite sign of $A_N$ at $x_{F}>0$ was shown in the previous measurements. The bands at $x_{F}>0.15$ are obtained from twist-3 model calculations~\cite{Gamberg:2017gle} with a variation of $K/\pi$ ratio by $\pm 30\%$ relative to the central value.

\begin{figure}[thb!]
\centering
\begin{subfigure}[b]{0.49\linewidth}
\includegraphics[width=1.0\linewidth]{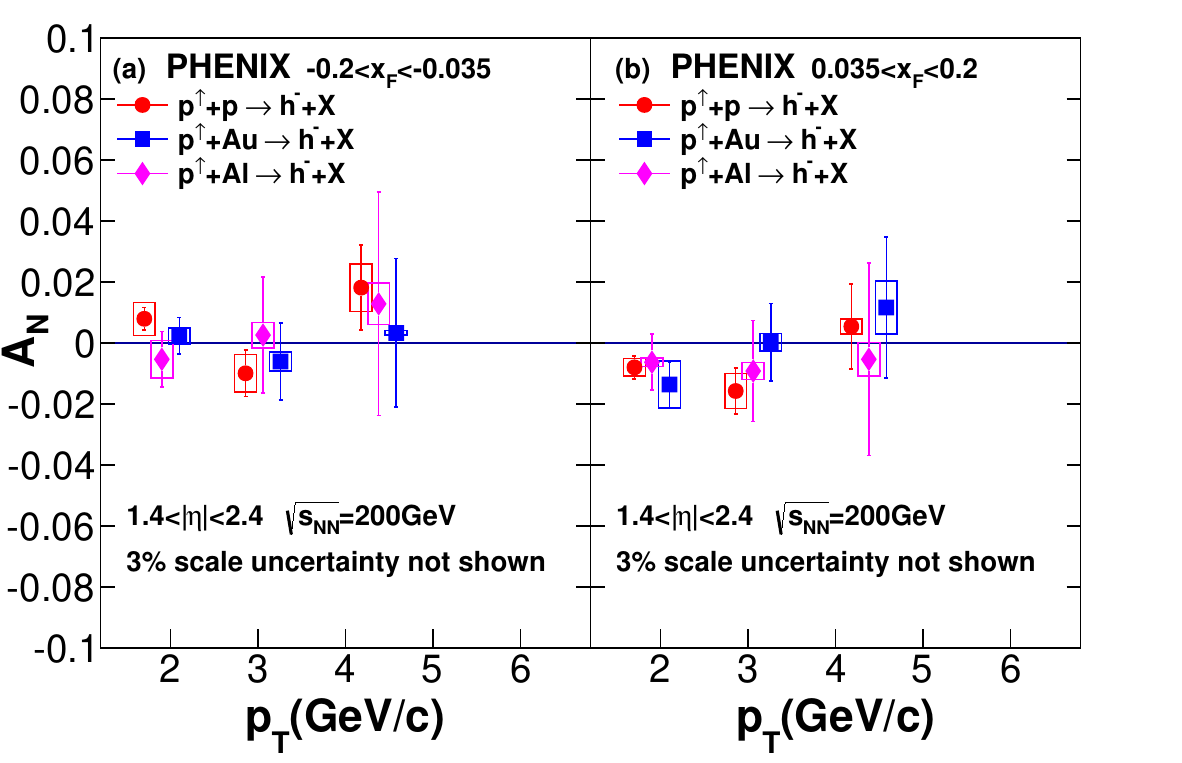}
\label{fig:AN_pt_neg}
\end{subfigure}
\begin{subfigure}[b]{0.49\linewidth}
\includegraphics[width=1.0\linewidth]{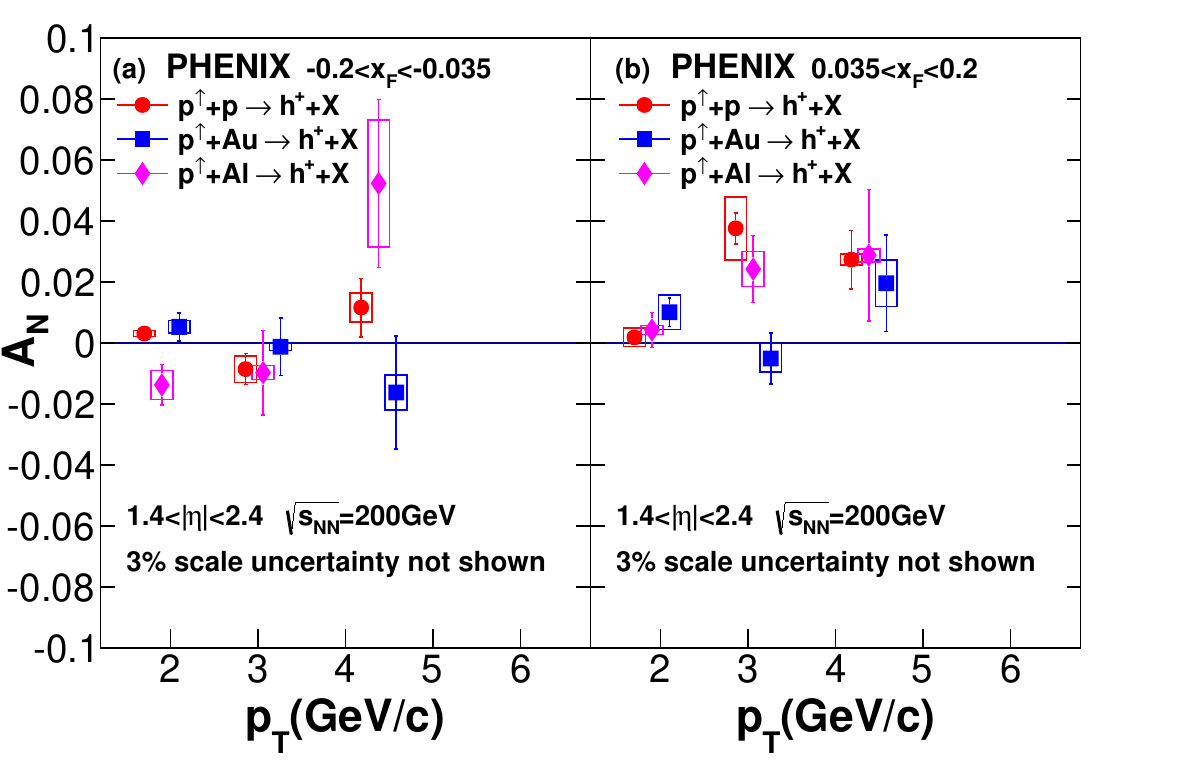}
\label{fig:AN_pt_pos}
\end{subfigure}
\caption{\label{fig:AN_pt_pA}
$A_N$ of negatively (Left) and positively-charged hadrons (Right) from \polpp, \polpAl, and \polpAu collisions as a function of $p_T$ in the backward ($x_F<0$) and forward ($x_F>0$) regions~\cite{PHENIX:2023axd}. Vertical bars (boxes) represent statistical (systematic) uncertainties.} 
\end{figure}

\begin{figure}[thb]
\centering
\includegraphics[width=1.0\linewidth]{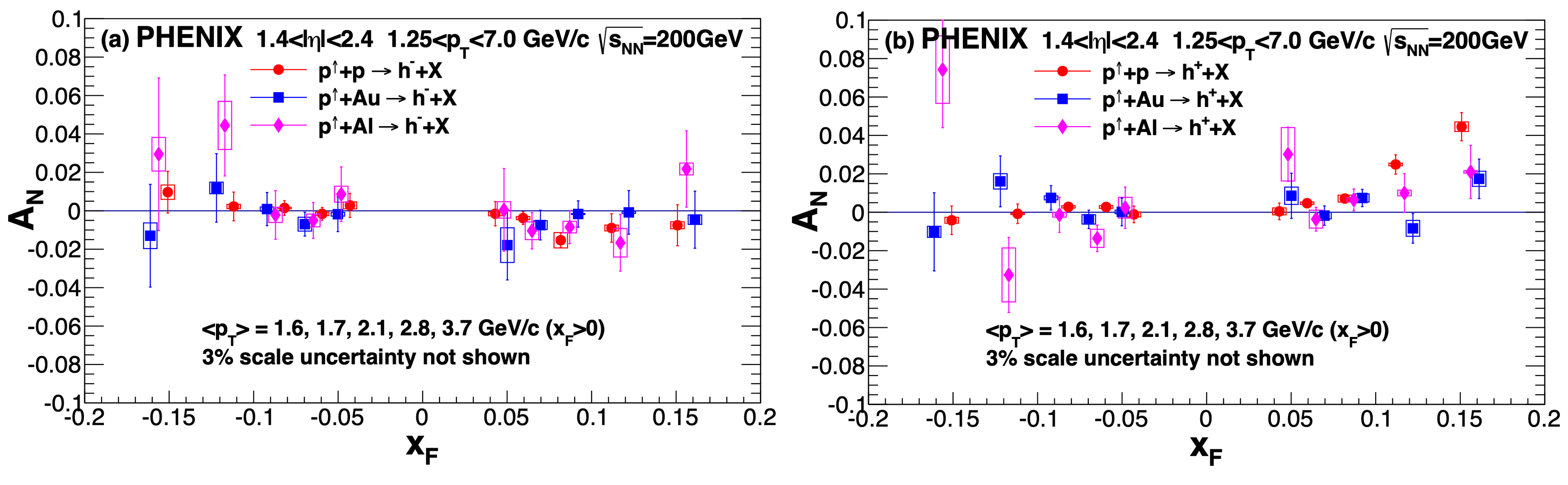}
\caption{\label{fig:AN_xf_pA}
$A_N$ of negatively (Left) and positively-charged hadrons (Right) from \polpp, \polpAl, and \polpAu collisions as a function of $x_F$, where $x_F>0$ is along the direction of the polarized proton~\cite{PHENIX:2023axd}. Vertical bars(boxes represent statistical (systematic) uncertainties.} 
\end{figure}

Fig.~\ref{fig:AN_pt_pA} and~\ref{fig:AN_xf_pA} show $A_N$ for negatively and positively charged hadrons in \polpp and \polpA collisions as a function of $p_T$ and $x_F$~\cite{PHENIX:2023axd}. $A_N$ for positively charged hadrons at large positive $x_F$ in \polpAl and \polpAu are smaller than the one in \polpp. For negatively charged hadrons at $x_{F}>0$, the deviation of $A_N$ in \polpp and \polpAu is not significant to state a modification due to large uncertainty.

\section{Summary}
In this study, the transverse single spin asymmetry ($A_N$) of positively and negatively charged hadrons ($h^{\pm}$) at forward and backward rapidity ($1.4<|\eta|<2.4$) over the range of $1.5<p_{T}<7.0 {\rm\ GeV}/c$ and $0.04<|x_{F}|<0.2$ from transversely polarized proton-proton (\polpp) and proton-nucleus (\polpAl, \polpAu) collisions are presented while the previous study is limited to positively charged hadrons at $0.1<x_{F}<0.2$~\cite{PHENIX:2019ouo}. 
The results at $x_{F}<0$ are consistent with zero in all collision systems. At $x_{F}>0$, negative charged hadron results show small negative to zero $A_{N}$ in \polpp and \polpA collisions. $A_{N}$ for positively charged hadrons increases to positive values as $x_{F}$ increases at $x_F>0$ in \polpp collisions. However, \polpAu results show suppression of $A_{N}$ at $0.1<x_{F}<0.2$ compared to the \polpp result. The results will be useful to understand the origin of $A_{N}$ and offer a tool to investigate phenomena and nuclear effects in small-$x$. 

\bibliographystyle{utphys} 
\bibliography{DIS23_jbok}

\providecommand{\href}[2]{#2}\begingroup\raggedright\begin{thebibliography}{10}

\bibitem{Aschenauer:2016our}
E.-C. Aschenauer {\em et~al.}, ``{The RHIC Cold QCD Plan for 2017 to 2023: A
  Portal to the EIC},'' \href{https://arxiv.org/abs/1602.03922}{{\ttfamily
  arXiv:1602.03922 [nucl-ex]}}.

\bibitem{BRAHMS:2008doi}
{\bfseries BRAHMS} Collaboration, I.~Arsene {\em et~al.}, ``{Single Transverse
  Spin Asymmetries of Identified Charged Hadrons in Polarized $pp$ Collisions
  at $\sqrt{s}$ = 62.4 GeV},''
  \href{https://dx.doi.org/10.1103/PhysRevLett.101.042001}{{\em Phys. Rev.
  Lett.} {\bfseries 101} (2008) 042001}.

\bibitem{Kang:2011ni}
Z.~Kang and F.~Yuan, ``{Single-Spin Asymmetry Scaling in the Forward Rapidity
  Region at RHIC},'' \href{https://dx.doi.org/10.1103/PhysRevD.84.034019}{{\em
  Phys. Rev. D} {\bfseries 84} (2011) 034019}.

\bibitem{Kovchegov:2012ga}
Y.~V. Kovchegov and M.~D. Sievert, ``{A New Mechanism for Generating a Single
  Transverse Spin Asymmetry},''
  \href{https://dx.doi.org/10.1103/PhysRevD.86.034028}{{\em Phys. Rev. D}
  {\bfseries 86} (2012) 034028}. [(E). Phys.Rev.D 86, 079906 (2012)].

\bibitem{Hatta:2016wjz}
Y.~Hatta, B.~Xiao, S.~Yoshida, and F.~Yuan, ``{Single-Spin Asymmetry in Forward
  $pA$ Collisions},'' \href{https://dx.doi.org/10.1103/PhysRevD.94.054013}{{\em
  Phys. Rev. D} {\bfseries 94} no.~5, (2016) 054013}.

\bibitem{Hatta:2016khv}
Y.~Hatta, B.~Xiao, S.~Yoshida, and F.~Yuan, ``{Single-spin asymmetry in forward
  $pA$ collisions II: Fragmentation contribution},''
  \href{https://dx.doi.org/10.1103/PhysRevD.95.014008}{{\em Phys. Rev. D}
  {\bfseries 95} no.~1, (2017) 014008}.

\bibitem{Benic:2018amn}
S.~Beni\'c and Y.~Hatta, ``{Single-spin asymmetry in forward $pA$ collisions:
  Phenomenology at RHIC},''
  \href{https://dx.doi.org/10.1103/PhysRevD.99.094012}{{\em Phys. Rev. D}
  {\bfseries 99} no.~9, (2019) 094012}.

\bibitem{PHENIX:2019gix}
{\bfseries PHENIX} Collaboration, C.~Aidala {\em et~al.},
  ``{Nuclear-modification factor of charged hadrons at forward and backward
  rapidity in $p+$Al and $p+$Au collisions at $\sqrt{s_{_{NN}}}=200$ GeV},''
  \href{https://dx.doi.org/10.1103/PhysRevC.101.034910}{{\em Phys. Rev. C}
  {\bfseries 101} no.~3, (2020) 034910}.

\bibitem{PHENIX:2017wbv}
{\bfseries PHENIX} Collaboration, C.~Aidala {\em et~al.}, ``{Cross section and
  transverse single-spin asymmetry of muons from open heavy-flavor decays in
  polarized $p$+$p$ collisions at $\sqrt{s}=200$ GeV},''
  \href{https://dx.doi.org/10.1103/PhysRevD.95.112001}{{\em Phys. Rev. D}
  {\bfseries 95} no.~11, (2017) 112001}.

\bibitem{PHENIX:2019ouo}
{\bfseries PHENIX} Collaboration, C.~Aidala {\em et~al.}, ``{Nuclear Dependence
  of the Transverse Single-Spin Asymmetry in the Production of Charged Hadrons
  at Forward Rapidity in Polarized $p+p$, $p+$Al, and $p+$Au Collisions at
  $\sqrt{s_{_{NN}}}=200$ GeV},''
  \href{https://dx.doi.org/10.1103/PhysRevLett.123.122001}{{\em Phys. Rev.
  Lett.} {\bfseries 123} no.~12, (2019) 122001}.

\bibitem{PHENIX:2012itj}
{\bfseries PHENIX} Collaboration, A.~Adare {\em et~al.},
  ``{Nuclear-Modification Factor for Open-Heavy-Flavor Production at Forward
  Rapidity in Cu+Cu Collisions at $\sqrt{s_{NN}}=200$ GeV},''
  \href{https://dx.doi.org/10.1103/PhysRevC.86.024909}{{\em Phys. Rev. C}
  {\bfseries 86} (2012) 024909}.

\bibitem{PHENIX:2013txu}
{\bfseries PHENIX} Collaboration, A.~Adare {\em et~al.}, ``{Cold-Nuclear-Matter
  Effects on Heavy-Quark Production at Forward and Backward Rapidity in d+Au
  Collisions at $\sqrt{s_{NN}}=200$ GeV},''
  \href{https://dx.doi.org/10.1103/PhysRevLett.112.252301}{{\em Phys. Rev.
  Lett.} {\bfseries 112} no.~25, (2014) 252301}.

\bibitem{PHENIX:2023axd}
{\bfseries PHENIX} Collaboration, N.~J. Abdulameer {\em et~al.}, ``{Transverse
  single-spin asymmetry of charged hadrons at forward and backward rapidity in
  polarized $p$+$p$, $p$+Al, and $p$+Au collisions at $\sqrt{s_{NN}}=200$
  GeV},'' \href{https://arxiv.org/abs/2303.07191}{{\ttfamily arXiv:2303.07191
  [hep-ex]}}.

\bibitem{Gamberg:2017gle}
L.~Gamberg, Z.~Kang, D.~Pitonyak, and A.~Prokudin, ``{Phenomenological
  constraints on $A_N$ in $p^\uparrow p\to \pi\, X$ from Lorentz invariance
  relations},'' \href{https://dx.doi.org/10.1016/j.physletb.2017.04.061}{{\em
  Phys. Lett. B} {\bfseries 770} (2017) 242}.

\end{thebibliography}\endgroup

\end{document}